\documentclass[11pt]{article}
\usepackage{amsmath,amsfonts,amssymb,latexsym}

\newtheorem{theorem}{Theorem}[section]

\def\be{\begin{equation}}
\def\ee{\end{equation}}
\def\bq{\begin{eqnarray}}
\def\eq{\end{eqnarray}}
\def\beq{\begin{eqnarray*}}
\def\eeq{\end{eqnarray*}}

\begin{document}
\title{\Huge{Generic regular universes in higher order gravity theories}}
\author{\Large{ Spiros Cotsakis\footnote{\texttt{email:\thinspace
skot@aegean.gr}}, Dimitrios Trachilis\footnote{\texttt{email:\thinspace dtrachilis@aegean.gr}}\,
and Antonios Tsokaros\footnote{\texttt{email:\thinspace atsok@aegean.gr}}} \\
{\normalsize Research Group of Geometry, Dynamical Systems and
Cosmology}\\  {\normalsize University
of the Aegean}\\ {\normalsize Karlovassi 83 200, Samos,
Greece}}
\maketitle
\begin{abstract}
\noindent We review recent results on the  Cauchy-Kowalevsky structure of theories with higher derivatives in vacuum. We prove genericity of regularity of solutions under the assumption of analyticity. Our approach is framed  in the general context of formal series expansions of the metric around a regular point.
\end{abstract}

\noindent
In Ref. \cite{trachilis1}, we  treated the problem of the existence of generic perturbations of the regular state in higher order gravity theories in vacuum that derive from the lagrangian $R+\epsilon R^2$ under the assumption of analyticity. The field equations of this theory in vacuum are
\begin{equation}
L_{ij}=f'(R)R_{ij} -\frac{1}{2}f(R)g_{ij} -\nabla_i\nabla_j f'(R) +g_{ij}\Box f'(R)=0,
\label{eq:FEs}
\end{equation}
where we take $f(R)=R+\epsilon R^2$, $\epsilon\neq0$.  Here Greek suffixes go from 1 to 3,
while Latin ones from 0 to 3.
  We have shown that there exists a  first order formulation of the theory with the Cauchy-Kowalevsky property, similar to the one known in general relativity, cf. Refs. \cite{ycb}-\cite{ren2}. In this formulation, the field equations (\ref{eq:FEs}) become equivalent to the following system of evolution equations:
\begin{eqnarray}
\partial_t \gamma_{\alpha\beta} &=& -K_{\alpha\beta},
\label{eq:pg}\\
\partial_t K_{\alpha\beta} &=& D_{\alpha\beta},
\label{eq:pK}\\
\partial_t D_{\alpha\beta} &=& W_{\alpha\beta},
\label{eq:pD}\\
\partial_t W &=&\frac{1}{6\epsilon}( \frac{1}{2} P +\frac{1}{8} K^2  -\frac{5}{8} K^{\alpha\beta} K_{\alpha\beta} +D ) + \nonumber \\
&&\frac{1}{6}[P^2 +\frac{1}{4}P K^2 -\frac{1}{4}P K^{\alpha\beta} K_{\alpha\beta} +\frac{1}{32} K^4 -\frac{1}{16}K^2 K^{\alpha\beta} K_{\alpha\beta}  -\nonumber \\
&&6K K^\alpha_\beta K^\beta_\gamma K^\gamma_\alpha -\frac{99}{32} (K^{\alpha\beta} K_{\alpha\beta})^2 +27 K^\alpha_\beta K^\beta_\gamma K^\gamma_\delta K^\delta_\alpha +9 K K^{\alpha\beta} D_{\alpha\beta}  -  \nonumber \\
&&57 K^\alpha_\beta K^\beta_\gamma D^\gamma_\alpha  +\frac{13}{2}D K^{\alpha\beta} K_{\alpha\beta} -\frac{7}{2}D^2 +15 D^{\alpha\beta} D_{\alpha\beta}  -3K W   +\nonumber \\
&&15K^{\alpha\beta} W_{\alpha\beta}  -6\partial_t (\partial_t P) -\nonumber \\
&&4\gamma^{\alpha\beta} \nabla_\alpha \nabla_\beta (-P -D +\frac{3}{4} K^{\alpha\beta} K_{\alpha\beta} -\frac{1}{4}K^2)],
\label{eq:pW}
\end{eqnarray}
where the line element in a Cauchy adapted frame is given by $ds^2=dt^2-\gamma_{\alpha\beta}dx^\alpha dx^\beta, \gamma_{\alpha\beta}=-g_{\alpha\beta}$.  $K_{\alpha\beta}$ is the extrinsic curvature,  the \emph{acceleration tensor}  $D_{\alpha\beta}$ is introduced  through the \emph{second variational equation} (3), while we also introduced the \emph{jerk tensor} (3rd order derivatives) $W_{\alpha\beta}$ through the \emph{jerk} equation (4). Further, $P=\textrm{tr}P_{\alpha\beta}$ is the trace of the three dimensional Ricci tensor, and we find the existence of \emph{constraints}, analogous to the situation in general relativity,
\noindent
\emph{Hamiltonian Constraint}
\begin{eqnarray}
\mathcal{C}_0&=&\frac{1}{2}P  +\frac{1}{8}K^2 -\frac{1}{8}K^{\alpha\beta} K_{\alpha\beta} + \nonumber \\
&&\epsilon[-\frac{1}{2}P^2  -\frac{1}{4}P K^2 +\frac{1}{4}P K^{\alpha\beta} K_{\alpha\beta}  -\frac{1}{32}K^4 +\frac{1}{16}K^2 K^{\alpha\beta} K_{\alpha\beta} + \nonumber\\
&&\frac{3}{32}(K^{\alpha\beta} K_{\alpha\beta})^2 -\frac{1}{2}D K^{\alpha\beta} K_{\alpha\beta} +\frac{1}{2}D^2     - \nonumber\\
&&2 \gamma^{\alpha\beta} \nabla_\alpha \nabla_\beta(-P - \frac{1}{4}K^2 +\frac{3}{4}K^{\gamma\delta} K_{\gamma\delta}  -D)] =0,
\label{eq:hamiltonian}
\end{eqnarray}
\emph{Momentum Constraint}
\begin{eqnarray}
\mathcal{C}_\alpha&=&\frac{1}{2} (\nabla_\beta K^\beta_\alpha - \nabla_\alpha K) + \nonumber \\
&&\epsilon[(-P  -\frac{1}{4}K^2 +\frac{3}{4}K^\delta_\gamma K^\gamma_\delta -D) (\nabla_\beta K^\beta_\alpha - \nabla_\alpha K)  - \nonumber \\
&&\nabla_\alpha (-2\partial_t P + K K^{\gamma\delta} K_{\gamma\delta} -3 K^\beta_\gamma K^\gamma_\delta K^\delta_\beta -K D +5K^{\gamma\delta} D_{\gamma\delta} -2W)]=0.
\label{eq:momentum}
\end{eqnarray}
The constraints show that the \emph{initial data} $(\gamma_{\alpha\beta},K_{\alpha\beta},D_{\alpha\beta},W)$ cannot be chosen arbitrarily in this problem, but must satisfy  the equations (\ref{eq:hamiltonian}) and (\ref{eq:momentum}) on each slice $\mathcal{M}_{t}$. From this formulation, our first result is contained on the following theorem (local Cauchy problem-analytic case):
 \begin{theorem}
 For the lapse and shift given by $N=1,\beta=0$, and if we prescribe analytic initial data $(\gamma_{\alpha\beta},K_{\alpha\beta},D_{\alpha\beta},W)$ on some initial slice $\mathcal{M}_{0}$, then there exists a neighborhood of $\mathcal{M}_{0}$ in $\mathbb{R}\times\mathcal{M}$ such that the evolution equations  (\ref{eq:pg}), (\ref{eq:pK}), (\ref{eq:pD}) and (\ref{eq:pW}) have an analytic solution in this neighborhood consistent with these data. This analytic solution is the development of  the prescribed initial data on  $\mathcal{M}_{0}$ if and only if these initial data satisfy the constraints.
  \end{theorem}
There are  19 relations, that is 18 by the three evolution equations   (\ref{eq:pg}), (\ref{eq:pK}), (\ref{eq:pD}) and one from (\ref{eq:pW}). But we have the freedom to perform 4 diffeomorphism changes and we also have the  4 constraints (\ref{eq:hamiltonian}) and (\ref{eq:momentum}). Hence, the vacuum theory has $19-4-4=11$ degrees of freedom. This in turn implies that any solution with 11 free functions has the same degree of generality with a general solution of the theory.

In Ref. \cite{trachilis1}, we further studied the following problem: Suppose that we have a regular formal series representation of the spatial metric of the form
\begin{equation}
\gamma_{\alpha\beta}= \gamma^{(0)}_{\alpha\beta} +\gamma^{(1)}_{\alpha\beta}\;t + \gamma^{(2)}_{\alpha\beta}\;t^2 + \gamma^{(3)}_{\alpha\beta}\;t^3 + \gamma^{(4)}_{\alpha\beta}\;t^4 + \cdots,
\label{eq:3dimmetric}
\end{equation}
where the $ \gamma^{(0)}_{\alpha\beta} , \gamma^{(1)}_{\alpha\beta} ,\gamma^{(2)}_{\alpha\beta} , \gamma^{(3)}_{\alpha\beta} , \gamma^{(4)}_{\alpha\beta},\cdots$ are functions of the space coordinates. Given data $a_{\alpha\beta}$, $b_{\alpha\beta}$, $c_{\alpha\beta}$, $d_{\alpha\beta}$, $e_{\alpha\beta}$, analytic functions of the space coordinates, such that the coefficients $\gamma^{(\mu)}_{\alpha\beta}, \mu=0,\cdots 4,$ are prescribed,
\begin{equation}\label{free}
 \gamma^{(0)}_{\alpha\beta}= a_{\alpha\beta}, \,\,  \gamma^{(1)}_{\alpha\beta}=b_{\alpha\beta}, \,\,   \gamma^{(2)}_{\alpha\beta}= c_{\alpha\beta}, \, \,   \gamma^{(3)}_{\alpha\beta}= d_{\alpha\beta}, \,\,
     \gamma^{(4)}_{\alpha\beta}= e_{\alpha\beta},
\end{equation}
how many of these data are truly independent when (\ref{eq:3dimmetric}) is taken to be a possible solution of the evolution equations  (\ref{eq:pg}), (\ref{eq:pK}), (\ref{eq:pD}) and (\ref{eq:pW}) together with the constraints  (\ref{eq:hamiltonian}) and (\ref{eq:momentum})?
For any tensor $X$, using the formal expansion  ($\ref{eq:3dimmetric}$), we can recursively calculate the coefficients in the expansion
\begin{equation}
X_{\alpha\beta}= X^{(0)}_{\alpha\beta} +X^{(1)}_{\alpha\beta}\;t + X^{(2)}_{\alpha\beta}\;t^2 + X^{(3)}_{\alpha\beta}\;t^3 + X^{(4)}_{\alpha\beta}\;t^4 + \cdots,
\label{x tensor}
\end{equation}
in particular we can write down a general iterated formula for the $n$-th order term, $X^{(n)}_{\alpha\beta}$. In Ref. \cite{trachilis1}, we were able to prove that the imposition of the field equations leads to 15 relations between the
 30 functions of the perturbation metric  (\ref{eq:3dimmetric}), that is we are left with 15 free functions. Taking into account the freedom we have in performing  4 diffeomorphism changes,  we finally conclude that there are in total 11 free functions in the solution  (\ref{eq:3dimmetric}). This means that  the regular analytic solution (\ref{eq:3dimmetric}) corresponds to a general solution of the problem. Put it differently, `regularity is a generic feature of the $R+\epsilon R^2$ theory in vacuum under the assumption of analyticity'. We are thus led to the following genericity result.
 \begin{theorem}
Let $a_{\alpha\beta}$ be a smooth Riemannian metric, $b_{\alpha\beta},c_{\alpha\beta},d_{\alpha\beta}$ and $e_{\alpha\beta}$  be symmetric smooth tensor fields which are traceless  with respect to the metric $a_{\alpha\beta}$, i.e., they satisfy $b=c=d=e=0$.  Then there exists a formal power series expansion solution of the vacuum higher order gravity equations of the form (\ref{eq:3dimmetric}) such that:
\begin{enumerate}
\item It is unique
\item The coefficients $\gamma^{(n)}_{\alpha\beta}$ are all smooth
\item It holds that $\gamma^{(0)}_{\alpha\beta}=a_{\alpha\beta}$ and $\gamma^{(1)}_{\alpha\beta}=b_{\alpha\beta},$ $\gamma^{(2)}_{\alpha\beta}=c_{\alpha\beta}$, $\gamma^{(3)}_{\alpha\beta}=d_{\alpha\beta}$ and  $\gamma^{(4)}_{\alpha\beta}=e_{\alpha\beta}$.
\end{enumerate}
\end{theorem}
In the course of the proof of this result, uniqueness followed because all coefficients were found recursively, while smoothness follows because in no step of the proof did we found it necessary to lower the $\mathcal{C}^\infty$ assumption. We also note that $b_{\alpha\beta}$ and $c_{\alpha\beta}$ are necessarily transverse with respect to $a_{\alpha\beta}$.

Using similar perturbation methods, we are currently interested in the genericity of various radiation solutions in a similar context. We shall report on these results elsewhere, cf. Ref. \cite{trachilis2}.

\end{document}